TITLE: A Mediation Analysis of the Relationship Between Land Use Regulation Stringency and Employment Dynamics

AUTHORS: Uche Oluku and Shaoming Cheng
Department of Public Policy and Administration
Florida International University

ABSTRACT:

The paper examines the effects of stringent land use regulations, measured using the Wharton Residential Land Use Regulatory Index (WRLURI), on employment growth during the period 2010-2020 in the Retail, Professional, and Information sectors across 878 local jurisdictions in the United States. All the local jurisdictions exist in both (2006 and 2018) waves of the WRLURI surveys and hence constitute a unique panel data. We apply a mediation analytical framework to decompose the direct and indirect effects of land use regulation stringency on sectoral employment growth and specialization. Our analysis suggests a fully mediated pattern in the relationship between excessive land use regulations and employment growth, with housing cost burden as the mediator. Specifically, a one standard deviation increase in the WRLURI index is associated with an approximate increase of 0.8 percentage point in the proportion of cost burdened renters. Relatedly, higher prevalence of cost-burdened renters has moderate adverse effects on employment growth in two sectors. A one percentage point increase in the proportion of cost burdened renters is associated with 0.04 and 0.017 percentage point decreases in the Professional and Information sectors, respectively.

**INTRODUCTION**

Policy makers, academics, advocacy groups and housing developers often contend that more stringent land use regulations impose significant costs on the economy. Hsieh and Moretti (2019) estimated that the spatial misallocation of labor, due to stringent land use regulations in three high-productivity cities, New York, San Jose, and San Francisco, reduced the aggregate growth of the U.S. economy by 36% between 1964 and 2009. They further contend that Gross Domestic Product (GDP) would otherwise be 9% higher as of 2009. Glaeser and Gyourko (2018) also estimated GDP to be 2% less annually because of excessive land use regulations. Yet another estimate indicates that excessive land use regulations may cost the U.S. economy as much as $1.95 trillion, i.e., 13.6% of annual GDP by impeding the efficient migration of labor to more productive regions in the country (Osman, 2020). While growth in GDP is not synonymous with employment growth, they are highly correlated[1] (Seyfried, 2011; Levine, 2013; Mandel & Liebens, 2019).

Despite a significant and long-established connection, the mechanism behind the relationship between stringent land use regulations and economic outcomes is unclear. Existing studies, such as, Hsieh and Moretti (2019) and Saks (2008) emphasize the housing supply constraints induced by excessive land use regulations and conclude that spatial dispersion of labor results mainly from spatial dispersion of housing prices. This paper extends the earlier findings by focusing on housing unaffordability measured by cost burden,[2] which transcends housing prices but connects housing prices with household income. Relative-income-adjusted housing prices, particularly when household income is stagnant and significantly, and disproportionately lags skyrocketing housing prices, will exacerbate housing unaffordability challenges, which in turn will likely constrain economic and employment growth. This paper also examines a mediated pathway, in which restrictive land use regulation influences housing unaffordability which in turn affects employment growth and specialization. The magnitude of the mediated effect of housing cost burden, an intervening variable, will also be assessed. Specifically, the mediated pathway presumes that stringent land use regulations constrain housing supply and exacerbate housing affordability challenges (Glaeser & Gyourko, 2003; Quigley & Rosenthal, 2005; Lin & Wachter, 2019), which in turn may hinder employment growth (Chakrabarti and Zhang, 2010; Zabel, 2012), or economic growth (Anthony 2022; Oluku & Cheng, 2021).

---

[1] Okun's law, named after Arthur Okun, a Yale economist who first proposed the direct and positive relationship between GDP and employment growth based on empirical observation of data in 1962 (Sharkey, 2022).
[2] Housing cost burden is a widely used indicator of housing affordability and serves as a proxy for housing affordability in this paper. A homeowner or renter household is considered cost-burdened if 30% or more of its gross annual income is spent on homeowner costs, or rents, respectively.

  The mediated pathway is built upon and connects two interrelated yet disjointed bodies of literature: one on the likely contributing factor of stringent land use regulations on housing unaffordability and the other on potential consequences of increasing housing cost burden on employment growth. Mediation analytical framework and technique (MacKinnon, Krull, and Lockwood, 2000) will be applied to explain the relationship between land use regulation stringency and employment growth, through the mediating variable of housing cost burden. The analysis of the mediating role of housing cost burden will not only advance scholarly understandings of the connection between land use regulations and employment dynamics but also the mechanism underlying the relationship. It will also provide empirical evidence for policy discussions and options in urban and regional planning, public management, and economic development.

Initially created by Gyourko, Saiz and Summers (2008), the Wharton Residential Land Use Regulatory Index (WRLURI, hereafter) is widely used to rank communities' land-use regulatory stringency. The original composite index consists of 11 subindexes of local residential regulatory regimes based on surveys of over 2,000 jurisdictions across the U.S. in 2006. An updated WRLURI consisting of 12 subindexes[3] was created by Gyourko, Hartly, and Krimmel (2021) using surveys of over 2,450 primary suburban communities nationwide conducted in 2018. We merged both the 2006 ("WRLURI2006") and 2018 ("WRLURI2018") indexes in the analysis and identified 878 localities that exist in both surveys. The extent of the jurisdictions that appear in both rounds of surveys are comparable to the estimates in Gyourko, Hartley, and Krimmel (2021) and Mleczko and Desmond (2023).

All 878 jurisdictions are matched with the American Community Survey (ACS) data for the analysis. Using the longitudinal dataset of the 878 local jurisdictions provides a good opportunity to examine temporal changes of local land use regulations and the impacts of such changes on housing cost burden and employment growth. In addition to the overall WRLURI indexes, we examine three sub-categorical indexes to see the correlations between the subindexes and housing cost burdens, as well as employment growth. The three subindexes are density restriction index (DRI), local zoning approval index (LZAI), and supply restrictions index (SRI). The Affordable Housing Index (AHI) is another pertinent subindex that may be correlated with housing cost burden. However, the AHI is only included in the WRLURI2018 index, but not in the WRLURI2006 index. Hence, we could not examine the AHI when both the WRLURI2006 and WRLURI2018 indexes were merged longitudinally.

---

[3] The 11 original subindexes were: Local Political Pressure; State Political Involvement; Court Involvement; Local Project Approval; Local Zoning Approval; Local Assembly; Supply Restrictions; Density Restriction; Open Space; Exactions; and Approval Delay. The Affordable Housing subindex was added in 2018.

Three two-digit industrial sectors, namely, Retail Trade (NAICS 44-45, "Retail"),[4] Professional, Scientific, and Technical Services (NAICS 54, "Professional"),[5] and Information (NAICS 51, "Information"),[6] are selected and examined to show the differential effects of restrictive land use regulations and increasing housing cost burdens on sectoral employment growth. An increase in the proportion of the workforce employed in the three industrial sectors signifies employment growth. The three sectors, Retail, Professional, and Information make up 27% of all business establishments, employ 21% of the workforce, and account for 23% of U.S. payrolls according to the 2019 Annual Statistics of U.S. Businesses (SUSB) (U.S. Census Bureau, 2022).

Retail, Professional and Information workers represent a microcosm of the U.S. workforce with households along all income spectrums and different levels of educational attainment. Professional and Information workers generally attain higher levels of education, have more advanced skills, and command higher wages than Retail workers. For instance, as of January 2023, average weekly wage was $713, $1,455, and $1,744 for workers in the Retail, Professional, and Information sectors, respectively [Bureau of Labor Statistics (BLS), 2023]. Given this wage disparity, we expect housing cost burdens amongst Retail, Professional and Information workers to vary significantly. Moreover, minorities and young adults make up a disproportionate share of the Retail workforce compared to the other two sectors (BLS, 2020; Oluku & Cheng, 2021) and have relatively higher housing cost burdens [Joint Center for

---

[4] The Retail sector includes establishments involved in retailing merchandise and those providing services incidental to the sale of merchandise: https://www.bls.gov/iag/tgs/iag44-45.htm
[5] The Professional, Scientific, and Technical Services sector is comprised of establishments (with a high degree of expertise and training) specialized in performing professional, scientific, and technical activities for others: https://www.bls.gov/iag/tgs/iag54.htm.
[6] The Information sector comprises establishments engaged in producing and distributing information and cultural products. They also provide the means for processing, transmitting and distributing these products: https://www.bls.gov/iag/tgs/iag51.htm.

Housing Studies of Harvard University (JCHS), 2018]. Hence, using these three sectors provides some insight on the effects of housing unaffordability on employment growth across all demographics.

**LITERATURE REVIEW**

The administration of land use regulations in the United States is a very acrimonious activity (Mayer & Somerville, 2000), involving a political process that varies significantly across jurisdictions[7] (Osman, 2020). Legal authority for land use regulation was generally under state governments before its delegation to local governments in the 1950s. Decisions regarding land use regulations are now highly decentralized and controlled by over 18,000 towns and cities nationwide (Evenson, Wheaton, Gyourko, Quigley, 2003). Meanwhile, there is little oversight of these towns and cities that independently create land use regulations (Quigley & Raphael, 2005). Hence, the types of land use regulations that exist and their stringency vary significantly across jurisdictions.

Developers often face a myriad of challenges involving land use regulations that constrain housing supply and inflate home prices, including density restrictions, zoning permits, and building codes, amongst others. While there is a dearth of literature on the direct relationship between land use regulations and employment growth (Saks, 2008; Kim & Hewings, 2013; Hsieh & Moretti, 2019), several studies find that more stringent land use regulations can constrain housing supply and increase home prices, thus exacerbating housing affordability challenges (Mayer & Somerville, 2000; Glaeser & Gyourko, 2003; Quigley & Raphael, 2005;

---

[7] According to the U.S. Census Bureau, as of 2012, there were 35,879 general purpose governments nationwide, including 16,360 towns and townships, 3,031 counties, and 19,519 municipal governments [(National League of Cities (NLC), 2022; Osman, 2020].

Quigley & Rosenthal, 2005; Ikeda & Washington, 2015; Gyourko & Malloy, 2015; Calder, 2017; Lin & Wachter, 2019; Landis & Reina, 2021). Other studies have revealed an adverse relationship between housing unaffordability and employment growth (Chakrabarti and Zhang, 2010; Osei & Winters, 2018; Zabel, 2012), or economic growth (Anthony, 2022; Oluku & Cheng, 2021).

Major corporations or prospective entrepreneurs are respectively more likely to expand operations or establish new businesses in regions where housing is more affordable to their workforce, than in metro areas with limited availability of affordable housing (Landis & Reina, 2021; Oluku & Cheng, 2021; Wardrip et al., 2011). Hence, the housing sector can serve as a springboard for regional economic growth (Saks, 2008; Wardrip, Williams & Hague, 2011; Westover, 2015; and Schwartz, 2016). Growth in GDP clearly has a positive and significant effect on employment growth (Seyfried, 2011). However, employment growth is often regarded as a lagging economic indicator because a sustained decline in the unemployment rate usually occurs after other measures of economic growth have emerged (Levine, 2013).

For the average American household, housing remains the largest expenditure, rising from 33% of total household spending in 2017-2019, to 35% in 2020. Housing expenditure increased by 3.6% and 5.6% from 2019 to 2020, and from 2020 to 2021, respectively, and by 2021, housing expenditure was 34% of total household spending (BLS, 2022). In 2019, 16.7 million homeowners, 20% of all U.S homeowners, and 20.4 million renters, 46% of all U.S. renters were cost-burdened (JCHS, 2021). As of 2016, Florida and California were the states with the largest share of cost-burdened renters at 54.1% and 53.9%, respectively, while Alaska and Montana, at 37.1%, had the lowest share of cost-burdened renters (JCHS, 2017). Meanwhile, there are stark racial and ethnic disparities in rental cost burdens. In 2019, the cost-burdened proportion of Black and Hispanic renters was 54% and 52%, respectively, compared to 42% for White and Asian renters (JCHS, 2021).

To provide additional insight regarding the mediating role of housing cost burden in linking land use regulation stringency with economic outcomes, we review three categories of existing literature on the relationships between: i) land use regulations and employment growth (Saks, 2008; Kim and Hewings, 2013; Hsieh & Moretti, 2019); ii) land use regulations and housing unaffordability (Glaeser & Gyourko, 2003; Lin & Wachter, 2019; Landis & Reina, 2021; and iii) housing unaffordability and employment growth (Chakrabarti & Zhang, 2010; Zabel, 2012; Osei & Winters, 2018), or economic growth (Anthony 2022; Oluku & Cheng, 2021).

*Relationship between Land Use Regulations and Employment Growth:*

Saks (2008), Kim & Hewings (2013), and Hsieh & Moretti (2019) analyzed the relationship between land use regulations and employment growth in the United States. Saks (2008) investigated how land use regulations constrain housing supply and affects labor migration patterns and employment, while Kim & Hewings (2013) analyzed how variations in land use regulation stringency affects local housing supply, and how quickly households respond to employment opportunities through intraregional migration. Hsieh & Moretti (2019) examined how the spatial misallocation of labor in high productivity, high wage cities with excessive land use regulations results in limited employment growth and loss in economic output.

Saks (2008) used existing Census Bureau estimates of housing stock values in 2001 and 2002, and self-estimates of housing stock values in 1980, 1990 and 2000 based on data on building permits for residential construction to evaluate how variations in housing supply affects housing and labor market dynamics across the United States. Saks created an Index of Housing

Supply Regulation to measure land use regulation stringency in 82 Metropolitan Statistical Areas (MSAs) nationwide from an average of six different surveys, including the WRLURI.

Kim & Hewings (2013) used an exploratory correlation technique to examine the relationships between land use regulation stringency, the correlation between population and employment changes using 1990 and 2000 census-tract level data, and changes in mean commuting time in 40 large MSAs nationwide. In their study, Kim & Hewings (2013) used Saks' (2008) Index of Housing Supply Regulation to measure regulatory stringency. Using a spatial equilibrium model,[8] Hsieh & Moretti (2019) calculated the costs of the spatial misallocation of labor to the U.S. economy with 1964, 1965, 2008, and 2009 County Business Patterns (CBP) data, supplemented with 1960 and 1970 Census of Population, and the 2008 and 2009 ACS data across 220 MSAs nationwide.

Saks (2008), Kim & Hewings (2013), and Hsieh & Moretti (2019) all found evidence of an adverse relationship between excessive land use regulations and employment growth. Saks (2008) found that excessive regulation of housing supply increases the marginal cost of construction, decreases the elasticity of housing supply, and substantially affects employment and wage dynamics in locations with severely restrictive land use regulations. Saks also found that locations with less restrictive land use regulations had more residential construction. Saks observed that an increase in housing demand in markets with fewer regulations resulted in relatively smaller increases in home prices. Saks (2008) concluded that employment growth lagged in areas where housing supply is more constrained.

---

[8] Hsieh & Moretti (2019) "measures the aggregate productivity costs of local housing constraints through the prism of a Rosen-Roback model" (p. 20). In a spatial equilibrium, labor is mobile across cities and a productivity shock in one city affects wages and employment in other cities.

Kim & Hewings (2013) found that regions with high levels of regulatory restrictions "were more likely to show lower levels of correlation between tract-level population and employment changes, and with increasing mean commuting times between 1990 and 2000" (p. 20). They concluded that restrictive regulations could cause barriers to housing development and hinder the ability of households to relocate in response to job opportunities, thus resulting in longer commuting distances or greater spatial mismatches.

Hsieh & Moretti (2019) found that in housing-constrained cities, growth in productivity results in higher home prices and higher nominal wages rather than an increase in local employment. Essentially, in housing-constrained cities, the high cost of housing can impede access to such high productivity. Hsieh & Moretti (2019) estimated that more restrictive regulatory policies in three high productivity cities, New York, San Jose, and San Francisco constrained housing supply in those cities, and reduced the aggregate growth of the U.S. economy by 36% between 1964 and 2009, and that GDP would otherwise be 9% higher as of 2009. They concluded that productivity shocks due to misallocation of labor in New York, San Francisco and San Jose affected wages and labor productivity not only in those cities, but elsewhere.

***Relationship between Land Use Regulation and Housing Unaffordability:***

Glaeser & Gyourko, (2003), Lin & Wachter (2019), and Landis & Reina (2021) examined the relationship between excessive land use regulations and housing affordability. Glaeser & Gyourko, (2003) used MSA-level R. S. Means Company data[9] on construction costs, combined with the 1989 and 1999 American Housing Survey (AHS) data to evaluate housing affordability across the United States. Lin & Wachter (2019) used the WRLURI index to measure land use regulation stringency to assess its impact on home prices based on property transaction data in 185 California cities between 1993 and 2017. In contrast, Landis & Reina (2021) used multiple measures of regulatory stringency to analyze the effects of land use regulations on home prices and rents, between 2005 and 2016, in a more geographically expansive study—encompassing 336 MSAs—more than 85% of the MSAs nationwide.

---

[9] R.S. Means Company reports construction costs data for numerous American and Canadian cities per square foot of living area. The index can be used to compare costs across cities and over time.

Glaeser & Gyourko (2003) found that in most parts of the country, home prices were close to the cost of constructing a new home. However, in some areas with high home prices, particularly New York City and California, they found that restrictive zoning policies were highly correlated with high home prices. Glaeser & Gyourko (2003) concluded that all the evidence from their analysis suggests that zoning contributed to high home prices. In general, current trends indicate that rental cost burden tends to rise concurrently with home prices (Sanfilippo, 2021).

Like Glaeser & Gyourko (2003), Landis & Reina (2021) found that more restrictive land use regulations can raise home prices and rents, and that the relationship is stronger in local economies with higher productivity, or those experiencing job growth. However, Landis &Reina (2021) also found that the increase in median home values and rents was more strongly associated with local economies experiencing employment and payroll growth and less so with more highly regulated MSAs when compared to less regulated ones. They concluded that housing could become more affordable with less restrictive land use regulations in some high-cost markets in various MSAs particularly in California and Florida.

Lin & Wachter (2019) distinguished between three paths through which land use regulations can affect home prices. Two of these channels are the forces of demand and supply, which they categorized as the local effects, and the third channel is the household location choice, also known as the general equilibrium (GE) effect. Lin & Wachter's (2019) stated that the supply and demand effects were equal to "(4.38% vs 0.32% if referenced to the average city in California) and is heterogeneous across cities. The relationship still holds, even when the GE effects are included in the two channels (3.24% vs 0.27%)" (p. 1). Land use regulations had the greatest effect on home prices in Los Angeles. Lin & Wachter estimated that home prices in Los Angeles could drop by approximately 25% if the city's land use regulations were comparable to the least regulated cities in California.

*Relationship between Housing Affordability and Employment or Economic Growth:*

Chakrabarti & Zhang (2010), Zabel (2012), and Osei & Winters (2018) examined the relationship between housing unaffordability and employment growth, while Oluku and Cheng (2021), and Anthony (2022) examined the relationship between housing unaffordability and economic growth. Chakrabarti & Zhang (2010) used a two-stage least squares regression model to analyze 1980, 1990 and 2000 decennial census data across 317 MSAs and 3,146 counties in the United States to evaluate the relationship between housing unaffordability and regional employment growth. Additionally, they analyzed data for 29 California cities between 1993 and 2004.

Osei & Winters (2018) evaluated the effects of labor demand shocks on home prices during the periods 1995-2000, 2002-2007, 2002-2015, and 2010- 2015, and the variance across time in 321 MSAs using the Quarterly Census of Employment and Wages program data. Unlike Chakrabarti & Zhang (2010), Osei & Winters (2018) appear to assume reverse causality in the relationship between housing unaffordability and employment growth.

Zabel (2012) examined the role of housing markets in the migration of workers across 277 MSAs in the United States in response to employment shocks. Zabel used data on house prices, wages, employment, in- and out-migration, and housing permits from 1990 through 2006 to estimate a vector autoregressive (VAR) model to examine how these variables affect cross-city migration. Like Osei & Winters (2018), Zabel (2012) contends that workers relocate for employment opportunities in response to rising housing costs.

Chakrabarti & Zhang (2010) found an adverse relationship between housing unaffordability and employment growth. In their study, housing unaffordability is measured by the ratio of median housing price to median household income, where an increase denotes more housing unaffordability. Chakrabarti & Zhang's (2010) found that over a two-year period in the California cities, employment growth declined by two-percentage points with a one-unit increase in the housing unaffordability ratio. At the MSA and county levels nationwide, Chakrabarti & Zhang found that over a ten-year period, employment growth declined by 10 percentage points for a one-unit increase in the housing unaffordability ratio.

Osei & Winters (2018) found that labor demand shocks result in rising home prices. Specifically, Osei & Winters (2018) found that, "for the period 1995-2015, on average, a ten percent increase in the annual average employment level is associated with 9.6% increase in housing prices. Similarly, for the same time period, a ten percent increase in the total wage bill is associated with a 7.2% increase in housing prices" (p. 5). They concluded that housing market conditions can cause households to move for economic opportunities. Similarly, Zabel (2012) found that at the MSA level, people are willing to relocate for employment opportunities. Zabel also concluded that the propensity to relocate for jobs has helped the United States maintain a long-lasting competitive advantage.

Meanwhile, Oluku and Cheng (2021), and Anthony (2022) examined the relationship between housing unaffordability and economic growth. Like Chakrabarti & Zhang (2010), Oluku and Cheng (2021) analyzed panel data across 383 MSAs and 3,137 counties, from 2010 through 2016 using a system generalized method of moments (GMM) dynamic model to evaluate the relationship between housing unaffordability and the number of business establishments in three business sectors.

Anthony (2022) used change in per capita GDP over two periods, from 2000 to 2010 and from 2010 to 2015 as a measure of economic growth in his assessment of the impact of housing affordability on economic growth in 100 largest MSAs in the United States. To serve as a fixed-effects proxy of an MSA's regulatory environment, Anthony used the population density at a one-mile distance from the center of the MSA. In both the Oluku and Cheng's (2021), as well as the Anthony (2022) studies, a higher percentage of cost-burdened renters and homeowners signified more housing unaffordability.

Oluku and Cheng (2021) estimated that over a 7-year period, across 3,137 counties nationwide, if housing unaffordability increases by one percentage point annually, the number of Professional business establishments would shrink by 90,248. Specifically, Oluku and Cheng found that at the county level, for a one-percentage point increase in homeowner cost burden, the number of Retail, Information, and Professional business establishments declined by 0.612, 0.244, and 1.044 in one year, respectively. While a one-percentage point increase in renter cost burden resulted in a decrease in the number of Retail, Information, and Professional establishments by 0.300, 0.144, and 0.366 in one year, respectively.

Like Oluku and Cheng (2021), Anthony (2022) found an adverse relationship between housing unaffordability and economic growth. Anthony (2022) found that housing unaffordability was linked to negative economic growth in the 100 largest MSAs where two-thirds of the U.S. population resides and more than 60% of the GDP was generated. Anthony concluded that making housing more affordable nationwide was an economic imperative.

**RESEARCH QUESTIONS AND HYPOTHESES**

Land use regulation stringency is likely correlated with various unobserved social, economic, geographic, and environmental factors, so it would be challenging to use cross-sectional data to estimate its causal effects (Gyorko & Malloy, 2015). Thus, we utilize panel data on the proportion of the workforce employed in three industrial sectors across 878 local jurisdictions in the United States to examine the following four hypotheses on the effects of land use regulation stringency on employment growth between 2010 to 2020. All local jurisdictions exist in both rounds of WRLURI surveys, and thus constitute a unique longitudinal data. The first three hypotheses are consistent with the mediation analytical framework to decompose the direct and indirect pathways and examine whether housing unaffordability, the mediating variable, fully or partially mediates the effect of land use regulations on economic growth in the model, while the last and fourth hypothesis pertains to differentiated effects across industrial sectors. Graphically, the mediated relationships are depicted in Figure 1.

**Figure 1: Diagram of the Mediated Role of Housing Cost Burden**

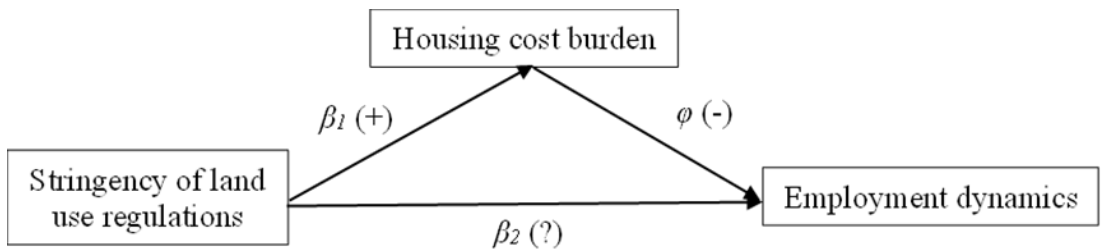

*Hypothesis 1: More stringent land use regulations are correlated with an increasing share of cost-burdened renters.*

*Hypothesis 2: An increase in the share of cost-burdened renters is correlated with slower employment growth.*

*Hypothesis 3: Housing cost burden serves as a mediator between land use regulations and employment growth.*

*Hypothesis 4: Additionally, we hypothesize that the mediating effect of housing unaffordability in the relationship between housing regulation stringency and economic growth may vary across industries.*

Compared to the Retail industry, which is somewhat ubiquitous across all jurisdictions, the Professional and Information industries are more footloose. The workforce in the Professional and Information sectors hence may be more likely affected by rising housing cost burden. Although firms in the Information and Professional sectors have relatively higher profit margins compared to their counterparts in the Retail sector (Damodaran, 2022) and may hedge against housing unaffordability challenges (Oluku & Cheng, 2021), employment growth and specialization in the two sectors may be disproportionately limited because of spatial misallocation of labor (Hsieh and Moretti, 2019). Specifically, as excessive land use regulations raise housing cost burden, housing unaffordability may become a greater hindrance to employment growth in the Professional and Information sectors. In contrast, in high cost, high income areas, though these places are likely to be cost burdened, employment in the Retail sector may be sustained in part because of the demand and purchasing power associated with high(er) income.

**MEDIATION MODEL**

Mediation analysis is generally applied to explain how two variables are related and to investigate the mediating variable or process underlying the observed relationship between an independent and an outcome variable. The mediating variable is an intermediate link between the independent and outcome variables. Specifically, the independent variable is related to the mediator, which in turn is related to the dependent variable (MacKinnon, 2013). Baron and Kenny (1986) laid out the steps to establish mediation, which have become the most widely used framework to test for mediation. The framework posits that the independent variable must affect the mediator and the mediator must affect the dependent variable when the independent variable is controlled. In addition, when the mediator is included and controlled, the independent and outcome variables must be related in a statistically insignificant manner. MacKinnon (2013) further contended that complete mediation is present when the independent variable no longer influences the dependent variable after the mediator has been controlled, while partial mediation occurs when the independent variable's influence on the dependent variable is reduced after the mediator is controlled.

Graphically, the mediated role of housing cost burden is shown in Figure 1, and mathematically, the mediation models can be expressed as:

$$Burden = \alpha_1 + \beta_1 Regulation + \delta_1 Control + \varepsilon_1 \qquad (1)$$

$$Emp = \alpha_2 + \beta_2 Regulation + \varphi Burden + \delta_2 Control + \varepsilon_2 \qquad (2)$$

In the mediation model, equation 1, the relationship between stringency of land use regulations (*Regulation*) and housing cost burden (*Burden*) is initially tested. Subsequently, in the outcome model, equation 2, the relationship of *Regulation* and *Burden* with the outcome variable employment growth (*Emp*) is tested. *Control* is a vector of control variables which are included in both mediation and outcome equations. The symbols $α_1$ and $α_2$ denote the intercepts; $β_1$, $β_2$, $δ_1$, $δ_2$, and $φ$ refer to the coefficients to be estimated; $ε_1$ and $ε_2$ are the error terms. According to Baron and Kenny (1986), to establish a mediation relationship, coefficient $β_1$ and $φ$ must be statistically significant, while coefficient $β_2$, in equation 2 when the mediating variable is added, must be statistically insignificant.

**DATA AND VARIABLE MEASUREMENT**

Housing cost burden is measured from renters' perspective, i.e., when a household spends 30% or more of its gross income on rent, the household is cost-burdened. The primary and exclusive focus on rental affordability is because renters are significantly more cost-burdened than homeowners (Oluku and Cheng, 2021) and rental cost burdens disproportionately affect low-income families (Larrimore & Schuetz, 2017; Gent & Leather, 2022; Mateyka & Yoo, 2023). Homeowners tend to have higher median incomes (Mateyka & Yoo, 2023) and are more insulated from rising housing costs because the majority of mortgages last 15-30 years, much longer than rental leases. Moreover, homeowners frequently refinanced their mortgages since 2008 to take advantage of historic low interest rates, but renters rarely benefit from such interest rate savings. Cost-burdened, often low-income households are unlikely to become homeowners in part because of difficulty in saving for down payments (Sanfilippo, 2021). So, they may be "stuck" in the rental market, struggling with increasing rents and unable to build wealth through homeownership.

The outcome variable, employment of localities that exist in both rounds of surveys, is measured respectively by sectoral employment in three selected industries in 2010 and 2020. The two decennial census years are selected because, in such decennial years, comprehensive data are more likely to be available at fine geographic levels. In addition, absolute sectoral employment is adjusted by total employment to control size variations across communities. The size-adjusted employment of the three selected industries provides insights not only on employment growth but also on changes in local employment specialization over time.

The WRLURI is used to measure the stringency of each community's local land-use regulatory environment. Two waves of WRLURI indexes are utilized for localities in both surveys. Most of the survey responses are 12 to 14 years apart, as the first WRLURI survey was sent out in late 2004, with the last round of responses received as late as 2006. The second survey was conducted entirely within the calendar year 2018. Because altering land use ordinances and regulations involves a formal, often lengthy and time consuming, legal process, the local regulatory regimes tend to be stable over a short period of time. However, the regulatory regime would be more likely to change over the longer time interval between the 2006 and 2018 surveys.

The aggregate measure—WRLURI—is comprised of 11 subindexes[10] that summarize information on the different aspects of the regulatory environment. Nine measures pertain to local characteristics, while two reflect state court and state legislative/executive branch behavior. Each index is designed so that a low value indicates a less restrictive or more laissez faire approach to land use regulations. Factor analysis is used to create the aggregate index, which is then standardized so that the sample mean is zero and the standard deviation is equal to one.

We measure change in the local regulatory environment by comparing overall and subcategory indexes of localities that completed both rounds of surveys. Gyourko, Hartley, and

---

[10] The original WRLURI2006 index consists of 11 subindexes. However, with the addition of a new subindex, the Affordable Housing Index, the WRLURI2018 index consists of 12 subindexes. For the analysis, we merged both the 2006 and 2018 indexes; hence the aggregate measure, WRLURI consists of 11 subindexes.

Krimmel (2019) systemically compared distributions of responses reported based on all respondents against those of communities that answered both surveys and concluded that responses were quite similar. Gyourko, et al. (2019) further contended that the strong similarities suggest that there were no "strong selection effects in who answered the underlying questions in both surveys that could bias our conclusions" (p. 26).

Figures 2 and 3 depict the spatial distribution of land-use regulatory stringency of all localities in both rounds of surveys, based on the WRLURI2006 and WRLURI2018 indexes, respectively. Both indexes are standardized with a mean of zero and a standard deviation of one (Gyourko, et al., 2021). In both figures, the red color signifies jurisdictions whose land-use regulations are more stringent than the average, while blue represents jurisdictions whose land-use regulatory environment is less stringent than the average. Within the respective red or blue categories, the size of the symbols indicates relative severity of local land-use regulations. Specifically, larger red symbols demonstrate more stringency while larger blue symbols indicate less stringency. Both Figures 2 and 3 suggest that the states of Arizona, Colorado, and Florida; the New England region, the Midwest area, and the Pacific coastal states stand out as highly regulated places in both surveys of land-use regulatory environments. Compared to Figure 2, Figure 3 shows a more spatially dispersed pattern of highly regulated land use policies, suggesting more jurisdictions had tightened land-use policies since the first survey in 2006. In addition to clustered red dots shown in Figure 2, new red dots appear in Figure 3 primarily in Southern states, such as North Carolina, South Carolina, Tennessee, and Texas, as well as Missouri in the Midwestern region of the United States.

[Insert Figure 2 here]

[Insert Figure 3 here]

Figure 4 presents each jurisdiction's temporal change between the WRLURI2006 and WRLURI2018 indexes. Although both indexes are standardized with a mean of zero and a standard deviation of one, comparison of a jurisdiction's indexes may still shed light on its temporal shifting in local land-use regulations. The primary reason is because the average land-use regulatory severity seems to be stable and comparable over the decade (Gyourko, Hartley, and Krimmel, 2021). Gyourko et al. also contended that the overall local regulatory environment appeared to be somewhat more restrictive in WRLURI2018 than it was in WRLURI2006. The standardized indexes suggest relative positioning in relation to the averages. When the averages of the two surveys are comparable over time, it would be cautiously reasonable to compare WRLURI2006 and WRLURI2018 index scores for a given locality's temporal changes. If a jurisdiction's WRLURI2018 index score is higher than its WRLURI2006 index, this may indicate that its land use regulatory environment has become more stringent over the two rounds of surveys.

[Insert Figure 4 here]

In Figure 4, red dots represent localities whose relative positioning in land-use regulations has become more stringent over time. It is apparent that red dots are more widespread in Figure 4 than in Figures 2 and 3, thus indicating a greater number of jurisdictions whose relative, temporal positioning became more stringent over the two rounds of surveys. Tightening local land use regulations seems to be a general, national trend. However, there are clusters of blue dots in the New England and Midwest regions, suggesting likely easing-off of restrictive local land use policies between the two rounds of surveys. The magnitude of such regulatory easing-off, indicated by the sizes of the blue dots, is noticeably smaller than the extent of tightening up of land use policies, as the red dots are much larger and more prevalent in space.

Covariates are selected in line with earlier studies and according to stages of the mediation analysis, where the first stage includes a dependent variable (DV) of prevalence of cost burdened renters and the second stage has DVs of employment dynamics of selected industries. In the first stage, there are three categories of independent variables, namely, housing, demographic, and economic factors (please see Table 1). The housing factors include rental vacancy rate, median rent, percentage of low-income housing tax credit (LIHTC) units[11] among all rental units, and percentage of HUD-subsidized rental units[12] among all rental units. The demographic control variables consist of percentages of female, Black, and Hispanic populations, median age, and percentage of adult population who have a bachelor's degree or higher. The economic factors are comprised of median household income and unemployment rate of adult population, 16 years or older. In both stages, in addition to the abovementioned independent variables, a given industry's employment of the county where a locality resides is incorporated to control for economic, agglomeration, and/or spatial factors that may influence sectoral employment around cities of interest.

[Insert Table 1 here]

**EMPIRICAL RESULTS**

The two-stage mediation models are applied in the analyses and Table 2 presents empirical results of the overall WRLURI index and its effects on employment growth in the Professional, Information, and Retail sectors. For any given industry, at the first stage, the mediating variable, specifically, a local jurisdiction's percentage of cost-burdened renters (households spending 30% or more of gross income on rent), is the dependent variable. In all three first-stage models, cities' overall WRLURI scores are positively correlated, with statistical significance, with their percentages of cost-burdened renters, suggesting that more restrictive land use regulations are likely a contributing factor to higher proportions of cost-burdened renters. This confirms the first hypothesis and is consistent with prior findings linking excessive land use regulatory environment with housing unaffordability. It is estimated that one-unit increase of the WRLURI overall index score, i.e., one standard deviation increase—the WRLURI index is standardized with a mean of zero and a standard deviation of one—is associated with approximately 0.8 percentage point increase in the proportion of cost burdened renters. The estimates of the increase in the proportion of cost burdened renters are consistent, ranging from 0.748 for the Retail sector to 0.849 for the Information sector.

---

[11] HUD national database of LIHTC housing: LIHTC Database Access (huduser.gov)
[12] HUD national database of all assisted housing: Assisted Housing: National and Local | HUD USER

[Insert Table 2 here]

At the second stage, the employment variables of the three sectors (Professional, Information, and Retail) are regressed against the mediating variable and the independent variable, along with all control variables. In all second-stage models in Table 2, the mediating variable, proportion of cost-burdened renters, has a negative and statistically significant correlation, respectively, with the proportions of employment in both the Professional and Information sectors. Essentially, as cost-burdened renters become more prevalent, the proportion of the workforce employed in the Professional or Information sector is expected to decline. It is estimated that one percentage point increase in the proportion of cost burdened renters is respectively associated with 0.04 and 0.017 percentage point decreases in the Professional and Information sectors. The median change in proportion of cost burdened renters of all 878 localities between 2010 and 2020 is 1.70 percentage points.[13] An increase in the proportion of cost-burdened renters by the median amount is associated with 0.08 and 0.03 percentage points decreases in employment in the Professional and Information industries, respectively. In addition, the first and second quantiles of increases of cost burdened renters is 39.2 and 4.4 percentage points. Localities in the first quantile of cost burdened renters would experience a 1.57 percentage point and 0.67 percentage point decline respectively in the Professional and Information industries. Because employment proportions of the Professional and Information industries may be interpreted either as total-employment-adjusted, sector-specific employment sizes or as industrial specialization given an industry's relative share, the negative association between housing cost burden and employment proportions can be seen as decreased employment or declining specialization in the Professional or Information sectors. Either interpretation would highlight the detrimental effects of rising housing cost burden on employment growth and specialization.

In contrast to the Professional and Information sectors, the Retail industry does not have a statistically significant correlation with rental cost burden. The sectoral variation supports the hypothesis that the mediated relationship of land use regulations, via housing cost burden, on employment growth and specialization may vary across industries. This also reinforces findings of Oluku and Cheng (2021) that, while firms in the Professional and Information sectors have a higher profit margin which may help them better cope with rising housing cost burdens, employment of such industries is footloose and may hence be shifted subject to absolute and relative housing cost burdens. The implications on employment growth and specialization would be critical because these industries are essential to economic prosperity.

---

[13] The average change in proportion of cost burdened renters of all 878 localities in 2010-2020 is 2.02 percentage points, almost identical to the median change. This suggests a fairly symmetric distribution of changes in proportions of cost burdened renters over time.

Associated with the mediation analyses of all three industries, the overall WRLURI index has a statistically significant and positive correlation with the housing cost burden variable in the first-stage models, but does not have any statistically significant relationship with the sector employment variables in the second stage analyses. Collectively, the combination of statistical significance of the two-stage models indicates a full mediation pattern. In other words, the relationship between land use regulation stringency and local employment dynamics is fully mediated through the intervening variable, housing cost burden. The fully mediated pathway not only reinforces the general literature pertaining either to the relationships between land use regulation stringency and housing unaffordability, but also that between housing unaffordability and employment growth. It also specifically complements Hsieh and Moretti (2019) and Saks (2008) in that the hindrance, created by excessive land use regulations, to spatial labor allocation and employment growth may result not just from housing supply restrictions but also from rising housing cost burdens.

In addition, most covariates have expected signs. The prevalence of cost-burdened renters is negatively correlated with household median income but positively correlated with median rent. The findings are consistent with the conventional wisdom that higher household median income alleviates cost burden, while higher rent exacerbates housing unaffordability. Unemployment rate as well as demographic variables pertaining to a higher proportion of minority residents are also positively associated with housing cost burden, thus indicating that communities with a higher proportion of minorities are more likely to be cost-burdened. This is also consistent with similar findings of disparities in housing affordability indicating that Black and Hispanic households are saddled with higher housing cost burdens compared to Whites (JCHS, 2018; JCHS, 2021; Sanfilippo, 2021), a situation that worsened after the COVID-19 pandemic (JCHS 2021a). County-wide employment specialization of each of the three industries of interest is positively related to municipal employment growth. This, to a great extent, suggests that the county-level control variable likely captures unobserved regional characteristics that may influence local jurisdictions' employment.

Tables 3, 4, and 5 present empirical results of three WRLURI subindexes with the mediated analytic framework. The three subindexes are respectively, density restriction index (DRI), local zoning approval index (LZAI), and supply restrictions index (SRI). Regression results of DRI and SRI are highly consistent with those with the overall WRLURI index, suggesting a fully mediated pathway in which each subindex respectively influences housing cost burden which in turn affects employment in the Professional and Information industries. On the contrary, LZAI does not have a statistically significant relationship with housing cost burden, so local zoning approval procedures and/or ease may not be a determining factor affecting housing cost burden thus affecting local employment growth and specialization.

[Insert Table 3 here]

[Insert Table 4 here]

[Insert Table 5 here]

**CONCLUSIONS AND POLICY IMPLICATIONS**

In this study, we examined existing literature on three categories of covariant relationships, between: i) land use regulations and employment growth; ii) land use regulations and housing unaffordability; and iii) housing unaffordability and employment growth. We utilized panel data on the proportion of the workforce employed in three industrial sectors—

Retail, Professional, and Information—across 878 local jurisdictions in the United States to examine several hypotheses on the effects of land use regulation stringency on employment growth between 2010 to 2020. We hypothesized that more stringent land use regulations are correlated with higher proportions of cost-burdened renters and that a higher proportion of cost-burdened renters is adversely related to employment growth. We also theorized that the effects of housing cost burden, which is the intervening variable in the relationship between land use regulation and employment growth may vary across industries.

Land use regulation stringency is measured using the composite 2006 and 2018 WRLURI indexes and comprised of multiple subindexes. Three of the subindexes, namely the density restriction index (DRI), local zoning approval index (LZAI), and supply restrictions index (SRI) are also used as alternate measures of land use regulation stringency in the study. The proportion of cost-burdened renters in any jurisdiction serves as a proxy for housing unaffordability. In the study, an increase in the proportion of the workforce employed in three industrial sectors, Retail, Professional, and Information signifies employment growth. These three sectors make up 27% of all business establishments, employ 21% of the workforce, and account for 23% of U.S. payrolls. The workforce in the Retail, Professional, and Information sectors represents individuals across all demographics and every socio-economic status.

Applying a mediation analytical framework, we decomposed the direct and indirect pathways through which land use regulation affects employment growth, with housing cost burden as the mediating variable and reported several key findings. First, our analysis reveals that more restrictive land use regulations are likely a contributing factor to higher proportions of cost-burdened renters. This is consistent with existing literature, and corroborates our first hypothesis, that land use regulation stringency is positively correlated with a higher proportion of cost-burdened renters. Specifically, we found that a one standard deviation increase in the WRLURI index is associated with approximately 0.8 percentage point increase in the proportion of cost burdened renters (ranging from 0.748 to 0.849 percentage points in the Retail and Information sectors, respectively).

Secondly, we found that the mediating variable—the proportion of cost-burdened renters—has a negative and statistically significant relationship with the proportions of workers employed in two industrial sectors. Essentially, as the proportion of cost-burdened renters increases, the proportion of the workforce employed in the Professional or Information sector is expected to decline. Specifically, a one percentage point increase in the proportion of cost burdened renters is associated with 0.04 and 0.017 percentage point decrease in the Professional and Information sectors, respectively. However, the Retail sector does not have a statistically significant relationship with rental cost burden, which supports our hypothesis that the mediated relationship of land use regulations, via housing cost burden may vary across industries.

Thirdly, associated with the mediation analyses of all three industries, the overall WRLURI index has a statistically significant and positive correlation with the rental cost burden variable in the first-stage models, but does not have any statistically significant relationship with the sector employment variables in the second stage analyses, which is an indication that the relationship between land use regulation stringency and local employment dynamics is fully mediated through the intervening variable, housing cost burden. The fully mediated pathway also bolsters the credibility of previous studies with findings showing that relationships exist between land use regulation stringency and housing unaffordability, and between housing unaffordability and employment growth.

Finally, we found that communities with a higher proportion of Blacks and Hispanics have a higher proportion of cost-burdened renters, which reinforces the general literature that Black and Hispanic households nationwide have higher housing cost burdens compared to Whites (JCHS, 2018; JCHS, 2021; Sanfilippo, 2021). Meanwhile, the regression results using the DRI and SRI subindexes are highly consistent with those with the overall WRLURI index, suggesting a fully mediated pathway in which each subindex respectively influences housing cost burden which in turn affects employment in the Professional and Information industries. This is an indication that density and supply restrictions are two types of land use regulations that likely affect housing cost burden thus affecting local employment growth. Conversely, the LZAI subindex does not have a statistically significant relationship with housing cost burden, suggesting that local zoning approval procedures do not appear to affect housing cost burden.

    The findings provide empirical evidence for policy discussions by urban scholars, and have key implications for local public officials, and other decision-makers involved in the administration of land use regulations. Under certain circumstances, there are socio-economic justifications, environmental factors, and other acceptable rationalizations for land use regulations. However, the costs of regulations must not outweigh the public benefits. Meanwhile, our comparison of the WRLURI2006 and WRLURI2018 indexes shows that nationwide, land use regulations have become more widespread and more stringent over time. This is not very encouraging, as our findings underscore that excessive land use regulations can fuel the worsening of the housing affordability crisis, which in turn has the potential to impose economic costs and limit or reduce employment growth.

**Figure 2: WRLURI2006 index, red indicating above average**

**Figure 3: WRLURI2018 index, red indicating above average**

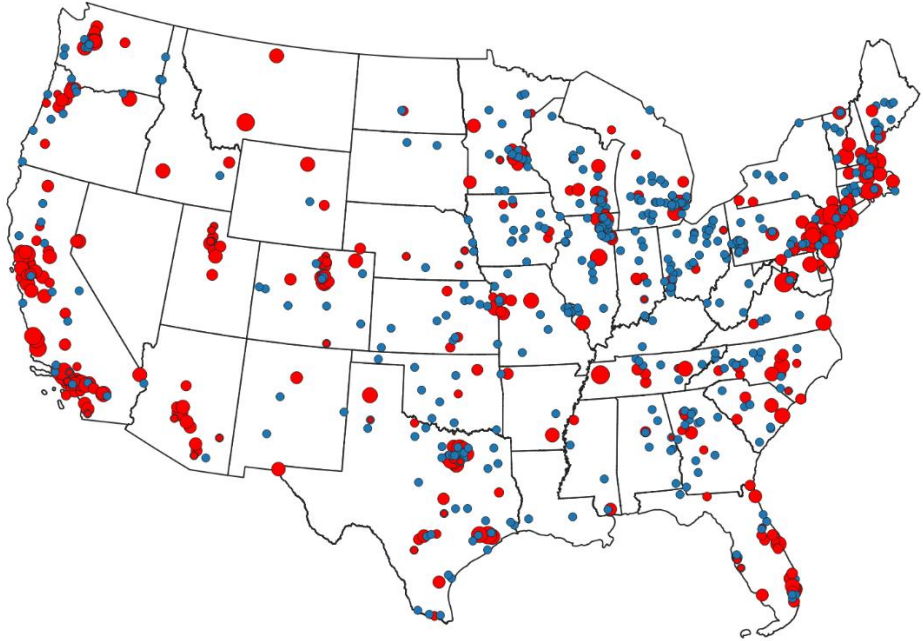

**Figure 4: Difference between two surveys**

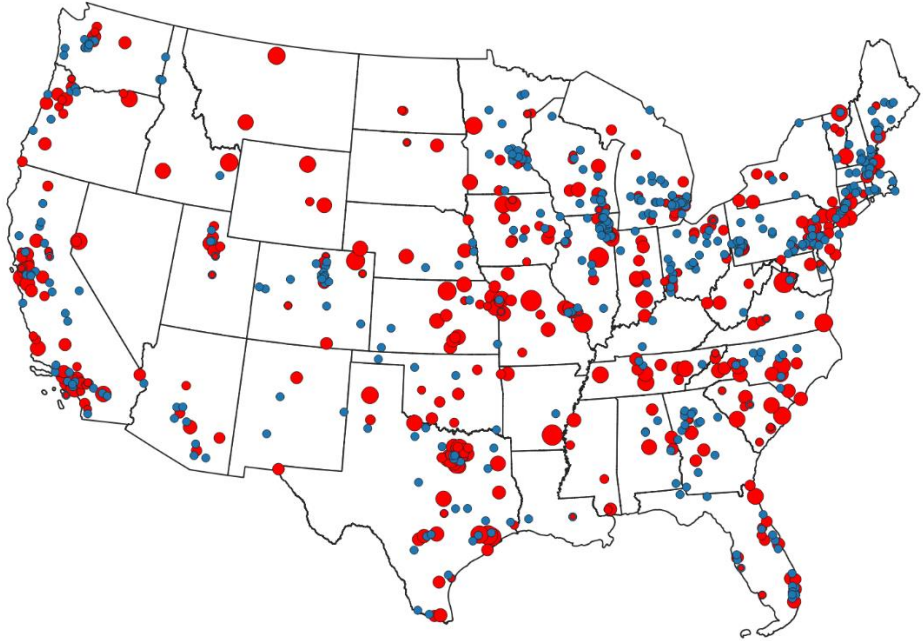

**Table 1: Variable Description and Measurement**

| Variables | Measurement | Sources |
|---|---|---|
| % of emp in professional | A locality's share of employment in the Professional sector in 2010 and 2020 | U.S. Census |
| % of emp in information | A locality's share of employment in the Information sector in 2010 and 2020 | |
| % of emp in retail | A locality's share of employment in the Retail sector in 2010 and 2020 | |
| WRLURI_overall | Overall WRLURI index | WRLURI |
| WRLURI_DRI | WRLURI density restriction index | |
| WRLURI_LZAI | WRLURI local zoning approval index | |
| WRLURI_SRI | WRLURI supply restrictions index | |
| % of cost burdened renters | Share of cost burdened renters who spend 30% or more gross income on rent | U.S. Census |
| Rental vacancy rate | A locality's rental vacancy rate | |
| Median HH income | A locality's median household income | |
| Median rent | A locality's median rent | |
| % of female pop | A locality's share of female population | |
| % of black pop | A locality's share of black population | |
| % of Hispanic pop | A locality's share of Hispanic population | |
| Median age | A locality's median age | |
| % of adult pop with bachelor's degree | A locality's share of adult population who have a bachelor's degree or higher | |
| Unemp rate | A locality's unemployment rate | |
| % of LIHTC unit | A locality's share of LIHTC units among all housing units | U.S. Census and HUD |
| % of HUD-assisted units | A locality's share of HUD-assisted units among all housing units | |
| County % of emp in professional | Share of employment in the Professional sector in the county where a locality is situated | U.S. Census |
| County % of emp in information | Share of employment in the Information sector in the county where a locality is situated | |
| County % of emp in retail | Share of employment in the Retail sector in the county where a locality is situated | |

## Table 2: Mediation Model Results with WRLURI Overall Index

| | Professional | | Information | | Retail | |
|---|---|---|---|---|---|---|
| | % of cost burdened renters | % of emp in professional | % of cost burdened renters | % of emp in information | % of cost burdened renters | % of emp in retail |
| % of cost burdened renters | | -0.040*** | | -0.017*** | | 0.010 |
| | | (0.010) | | (0.004) | | (0.008) |
| WRLURI_overall | 0.803** | 0.081 | 0.849** | 0.021 | 0.748* | 0.045 |
| | (0.399) | (0.101) | (0.398) | (0.045) | (0.398) | (0.083) |
| Rental vacancy rate | -0.131 | 0.060*** | -0.133 | -0.002 | -0.146* | -0.020 |
| | (0.086) | (0.023) | (0.084) | (0.007) | (0.086) | (0.019) |
| Median HH income | -0.000*** | 0.000*** | -0.000*** | -0.000 | -0.000*** | -0.000*** |
| | (0.000) | (0.000) | (0.000) | (0.000) | (0.000) | (0.000) |
| Median rent | 0.009*** | -0.000 | 0.010*** | 0.000** | 0.010*** | 0.001*** |
| | (0.002) | (0.000) | (0.002) | (0.000) | (0.002) | (0.000) |
| % of female pop | 0.175 | -0.064* | 0.169 | 0.008 | 0.170 | 0.021 |
| | (0.140) | (0.036) | (0.138) | (0.017) | (0.140) | (0.036) |
| % of black pop | 0.042** | 0.005 | 0.042** | -0.001 | 0.045** | -0.021*** |
| | (0.020) | (0.006) | (0.019) | (0.002) | (0.019) | (0.005) |
| % of Hispanic pop | 0.081*** | 0.013** | 0.086*** | -0.006** | 0.083*** | -0.025*** |
| | (0.016) | (0.006) | (0.016) | (0.003) | (0.016) | (0.006) |
| Median age | 0.046 | 0.054*** | 0.047 | -0.008 | 0.038 | -0.032* |
| | (0.061) | (0.016) | (0.061) | (0.007) | (0.060) | (0.019) |
| % of adult pop with bachelor's degree | 0.143*** | 0.106*** | 0.153*** | 0.016*** | 0.150*** | -0.058*** |
| | (0.032) | (0.008) | (0.031) | (0.003) | (0.031) | (0.006) |
| Unemp rate | 0.971*** | 0.092*** | 1.006*** | 0.030** | 0.956*** | -0.094*** |
| | (0.122) | (0.032) | (0.123) | (0.014) | (0.121) | (0.032) |
| % of LIHTC unit | -0.318 | 0.219 | -0.281 | 0.020 | -0.360 | 0.077 |
| | (0.650) | (0.218) | (0.631) | (0.075) | (0.627) | (0.181) |
| % of HUD-assisted units | -0.728** | -0.031 | -0.691** | -0.033 | -0.685* | 0.066 |
| | (0.351) | (0.096) | (0.352) | (0.038) | (0.351) | (0.098) |
| County % of emp in professional | -0.035 | 0.737*** | | | | |
| | (0.128) | (0.033) | | | | |
| County % of emp in information | | | -0.789** | 0.892*** | | |
|  |  |  | (0.376) | (0.061) |  |  |
| County % of emp in retail |  |  |  |  | 0.501** | 0.791*** |
|  |  |  |  |  | (0.195) | (0.065) |
| Constant | 31.118*** | 0.003 | 31.729*** | 0.168 | 25.252*** | 5.782*** |
|  | (7.218) | (1.894) | (7.110) | (0.885) | (7.289) | (2.045) |
| Observations | 1,653 | 1,653 | 1,653 | 1,653 | 1,653 | 1,653 |
| Adj R-squared | 0.260 | 0.788 | 0.264 | 0.493 | 0.264 | 0.429 |

Robust standard errors in parentheses, *** p<0.01, ** p<0.05, * p<0.1

## Table 3: Mediation Model Results with WRLURI Density Restriction Index (DRI) Sub-Index

|  | Professional | | Information | | Retail | |
|---|---|---|---|---|---|---|
|  | % of cost burdened renters | % of emp in professional | % of cost burdened renters | % of emp in information | % of cost burdened renters | % of emp in retail |
| % of cost burdened renters |  | -0.032*** |  | -0.013*** |  | 0.012 |
|  |  | (0.009) |  | (0.004) |  | (0.007) |
| WRLURI_DRI | 0.630** | -0.182** | 0.533** | -0.049 | 0.593** | 0.026 |
|  | (0.263) | (0.075) | (0.268) | (0.031) | (0.263) | (0.059) |
| Rental vacancy rate | -0.139 | 0.056** | -0.144* | -0.002 | -0.158* | -0.014 |
|  | (0.086) | (0.022) | (0.085) | (0.007) | (0.086) | (0.018) |
| Median HH income | -0.000*** | 0.000*** | -0.000*** | -0.000 | -0.000*** | -0.000*** |
|  | (0.000) | (0.000) | (0.000) | (0.000) | (0.000) | (0.000) |
| Median rent | 0.009*** | -0.000 | 0.010*** | 0.000** | 0.010*** | 0.001*** |
|  | (0.002) | (0.000) | (0.002) | (0.000) | (0.002) | (0.000) |
| % of female pop | 0.159 | -0.068* | 0.154 | 0.002 | 0.151 | 0.019 |
|  | (0.139) | (0.035) | (0.138) | (0.017) | (0.138) | (0.035) |
| % of black pop | 0.028 | 0.007 | 0.033* | -0.001 | 0.038** | -0.022*** |
|  | (0.020) | (0.006) | (0.019) | (0.002) | (0.019) | (0.005) |
| % of Hispanic pop | 0.083*** | 0.013** | 0.089*** | -0.007** | 0.090*** | -0.025*** |
|  | (0.016) | (0.006) | (0.016) | (0.003) | (0.016) | (0.006) |
| Median age | 0.078 | 0.059*** | 0.082 | -0.007 | 0.071 | -0.037* |
|  | (0.060) | (0.016) | (0.060) | (0.007) | (0.059) | (0.019) |
| % of adult pop with bachelor's degree | 0.158*** | 0.104*** | 0.165*** | 0.016*** | 0.171*** | -0.058*** |
|  | (0.030) | (0.008) | (0.030) | (0.003) | (0.030) | (0.006) |
| Unemp rate | 1.072*** | 0.061* | 1.074*** | 0.021 | 1.048*** | -0.097*** |
|  | (0.122) | (0.033) | (0.123) | (0.014) | (0.122) | (0.034) |
| % of LIHTC unit | -0.474 | 0.260 | -0.360 | 0.043 | -0.403 | 0.065 |
|  | (0.647) | (0.215) | (0.631) | (0.074) | (0.627) | (0.180) |
| % of HUD-assisted units | -0.794** | -0.036 | -0.780** | -0.030 | -0.737** | 0.079 |
|  | (0.343) | (0.093) | (0.344) | (0.037) | (0.345) | (0.094) |
| County % of emp in professional | 0.077 | 0.733*** |  |  |  |  |
|  | (0.127) | (0.032) |  |  |  |  |
| County % of emp in information |  |  | -0.339 | 0.863*** |  |  |
|  |  |  | (0.386) | (0.058) |  |  |
| County % of emp in retail |  |  |  |  | 0.589*** | 0.778*** |
|  |  |  |  |  | (0.196) | (0.064) |
| Constant | 29.387*** | -0.088 | 30.086*** | 0.398 | 23.094*** | 6.124*** |
|  | (7.074) | (1.861) | (7.002) | (0.899) | (7.129) | (2.003) |
|  |  |  |  |  |  |  |
| Observations | 1,722 | 1,722 | 1,722 | 1,722 | 1,722 | 1,722 |
| Adj R-squared | 0.273 | 0.789 | 0.273 | 0.489 | 0.277 | 0.439 |

Robust standard errors in parentheses, *** p<0.01, ** p<0.05, * p<0.1

## Table 4: Mediation Model Results with WRLURI Local Zoning Approval Index (LZAI) Sub-Index

|  | Professional | | Information | | Retail | |
|---|---|---|---|---|---|---|
|  | % of cost burdened renters | % of emp in professional | % of cost burdened renters | % of emp in information | % of cost burdened renters | % of emp in retail |
| % of cost burdened renters |  | -0.032*** |  | -0.013*** |  | 0.011 |
|  |  | (0.010) |  | (0.004) |  | (0.007) |
| WRLURI_LZAI | 0.140 | 0.013 | 0.074 | -0.012 | 0.129 | 0.019 |
|  | (0.270) | (0.070) | (0.270) | (0.035) | (0.271) | (0.059) |
| Rental vacancy rate | -0.146* | 0.058*** | -0.150* | -0.001 | -0.163* | -0.014 |
|  | (0.086) | (0.022) | (0.085) | (0.007) | (0.086) | (0.018) |
| Median HH income | -0.000*** | 0.000*** | -0.000*** | -0.000 | -0.000*** | -0.000*** |
|  | (0.000) | (0.000) | (0.000) | (0.000) | (0.000) | (0.000) |
| Median rent | 0.010*** | -0.000 | 0.010*** | 0.000** | 0.010*** | 0.001*** |
|  | (0.002) | (0.000) | (0.002) | (0.000) | (0.002) | (0.000) |
| % of female pop | 0.128 | -0.043 | 0.125 | 0.011 | 0.119 | 0.021 |
|  | (0.139) | (0.036) | (0.137) | (0.018) | (0.139) | (0.035) |
| % of black pop | 0.041** | 0.004 | 0.045** | -0.002 | 0.050*** | -0.022*** |
|  | (0.019) | (0.006) | (0.019) | (0.002) | (0.019) | (0.005) |
| % of Hispanic pop | 0.083*** | 0.012* | 0.091*** | -0.008*** | 0.089*** | -0.025*** |
|  | (0.016) | (0.006) | (0.016) | (0.003) | (0.016) | (0.006) |
| Median age | 0.075 | 0.061*** | 0.079 | -0.008 | 0.067 | -0.035* |
|  | (0.060) | (0.016) | (0.059) | (0.006) | (0.059) | (0.019) |
| % of adult pop with bachelor's degree | 0.155*** | 0.103*** | 0.166*** | 0.014*** | 0.167*** | -0.057*** |
|  | (0.030) | (0.009) | (0.029) | (0.004) | (0.030) | (0.006) |
| Unemp rate | 0.991*** | 0.091*** | 1.013*** | 0.026* | 0.969*** | -0.092*** |
|  | (0.120) | (0.031) | (0.121) | (0.014) | (0.119) | (0.032) |
| % of LIHTC unit | -0.482 | 0.190 | -0.393 | 0.015 | -0.433 | 0.097 |
|  | (0.635) | (0.209) | (0.621) | (0.073) | (0.617) | (0.175) |
| % of HUD-assisted units | -0.837** | -0.061 | -0.810** | -0.043 | -0.778** | 0.078 |
|  | (0.339) | (0.094) | (0.340) | (0.038) | (0.340) | (0.093) |
| County % of emp in professional | 0.059 | 0.740*** |  |  |  |  |
|  | (0.126) | (0.032) |  |  |  |  |
| County % of emp in information |  |  | -0.593 | 0.897*** |  |  |
| | | (0.375) | | (0.060) | | |
| County % of emp in retail | | | | | 0.581*** | 0.785*** |
| | | | | | (0.193) | (0.064) |
| Constant | 31.394*** | -1.741 | 32.130*** | -0.166 | 25.182*** | 5.825*** |
| | (7.040) | (1.988) | (6.918) | (0.954) | (7.116) | (1.980) |
| | | | | | | |
| Observations | 1,746 | 1,746 | 1,746 | 1,746 | 1,746 | 1,746 |
| Adj R-squared | 0.264 | 0.784 | 0.265 | 0.483 | 0.268 | 0.435 |

Robust standard errors in parentheses, *** p<0.01, ** p<0.05, * p<0.1

# Table 5: Mediation Model Results with WRLURI Supply Restrictions Index (SRI) Sub-Index

|  | Professional | | Information | | Retail | |
| --- | --- | --- | --- | --- | --- | --- |
|  | % of cost burdened renters | % of emp in professional | % of cost burdened renters | % of emp in information | % of cost burdened renters | % of emp in retail |
| % of cost burdened renters |  | -0.033*** |  | -0.014*** |  | 0.009 |
|  |  | (0.010) |  | (0.004) |  | (0.007) |
| WRLURI_SRI | 0.937*** | -0.045 | 0.929*** | 0.074* | 0.963*** | 0.017 |
|  | (0.345) | (0.114) | (0.347) | (0.039) | (0.340) | (0.082) |
| Rental vacancy rate | -0.154* | 0.058*** | -0.157* | -0.002 | -0.172** | -0.016 |
|  | (0.087) | (0.022) | (0.086) | (0.007) | (0.087) | (0.018) |
| Median HH income | -0.000*** | 0.000*** | -0.000*** | -0.000 | -0.000*** | -0.000*** |
|  | (0.000) | (0.000) | (0.000) | (0.000) | (0.000) | (0.000) |
| Median rent | 0.009*** | -0.000 | 0.010*** | 0.000** | 0.010*** | 0.001*** |
|  | (0.002) | (0.000) | (0.002) | (0.000) | (0.002) | (0.000) |
| % of female pop | 0.165 | -0.040 | 0.163 | 0.014 | 0.157 | 0.028 |
|  | (0.137) | (0.036) | (0.135) | (0.018) | (0.137) | (0.035) |
| % of black pop | 0.047** | 0.004 | 0.049*** | -0.001 | 0.054*** | -0.021*** |
|  | (0.019) | (0.006) | (0.019) | (0.002) | (0.019) | (0.005) |
| % of Hispanic pop | 0.084*** | 0.012* | 0.090*** | -0.007*** | 0.089*** | -0.025*** |
|  | (0.016) | (0.006) | (0.016) | (0.003) | (0.016) | (0.006) |
| Median age | 0.065 | 0.061*** | 0.068 | -0.008 | 0.055 | -0.036* |
|  | (0.059) | (0.016) | (0.059) | (0.006) | (0.058) | (0.019) |
| % of adult pop with bachelor's degree | 0.151*** | 0.102*** | 0.161*** | 0.014*** | 0.162*** | -0.058*** |
|  | (0.030) | (0.009) | (0.029) | (0.004) | (0.030) | (0.006) |
| Unemp rate | 0.984*** | 0.092*** | 1.012*** | 0.028** | 0.962*** | -0.094*** |
|  | (0.118) | (0.031) | (0.118) | (0.014) | (0.117) | (0.031) |
| % of LIHTC unit | -0.477 | 0.209 | -0.408 | 0.017 | -0.457 | 0.099 |
|  | (0.631) | (0.209) | (0.616) | (0.073) | (0.613) | (0.175) |
| % of HUD-assisted units | -0.886*** | -0.070 | -0.861** | -0.048 | -0.822** | 0.064 |
|  | (0.339) | (0.094) | (0.339) | (0.038) | (0.340) | (0.093) |
| County % of emp in professional | 0.026 | 0.735*** |  |  |  |  |
|  | (0.125) | (0.032) |  |  |  |  |
| County % of emp in information |  |  | -0.609 | 0.906*** |  |  |
|  |  | (0.372) |  | (0.059) |  |  |
| County % of emp in retail |  |  |  |  | 0.620*** | 0.792*** |
|  |  |  |  |  | (0.194) | (0.064) |
| Constant | 30.127*** | -1.843 | 30.573*** | -0.353 | 23.288*** | 5.591*** |
|  | (7.003) | (1.973) | (6.875) | (0.964) | (7.077) | (1.983) |
|  |  |  |  |  |  |  |
| Observations | 1,741 | 1,741 | 1,741 | 1,741 | 1,741 | 1,741 |
| Adj R-squared | 0.253 | 0.784 | 0.255 | 0.489 | 0.259 | 0.429 |

Robust standard errors in parentheses, *** p<0.01, ** p<0.05, * p<0.1

2022, from https://www.census.gov/library/publications/2021/demo/p60-273.html

US Census Bureau. (2022). 2019 SUSB Annual Data Tables by Establishment Industry. Retrieved May 4, 2022, from https://www.census.gov/data/tables/2019/econ/susb/2019-susb-annual.html

US Department of Agriculture (USDA). (2020). USDA Farm Labor - Released May 28, **2020.** Retrieved May 21, 2022, from https://www.nass.usda.gov/Publications/Todays_Reports/reports/fmla0520.pdf

Wardrip, K, Williams, L & Hague, S. (2011). The Role of Affordable Housing in Creating Jobs and Stimulating Local Economic Development: A Review of the Literature. Retrieved May 3, 2022, from at: https://providencehousing.org/wp-content/uploads/2014/03/Housing-and-Economic-Development-Report-2011.pdf

Westover, C. N. (2015). Affordable Housing is an Economic Development Benefit. *The National Law Review, Volume V, No 57.* Retrieved May 4, 2022, from https://www.natlawreview.com/article/affordable-housing-economic-development-benefit

Zabel J (2012). Migration, housing market, and labor market responses to employment shocks. *Journal of Urban Economics* 72: 267–284.
46